\documentclass[preprint,3p]{elsarticle}
\usepackage{lineno,hyperref}
\usepackage{subfigure}
\usepackage{bm}
\usepackage{overpic}
\usepackage{graphicx}
\usepackage{mathrsfs}
\usepackage{amsfonts}
\usepackage{amssymb}
\usepackage{indentfirst}
\usepackage{footmisc}
\usepackage{multirow}
\usepackage{url}
\usepackage{amsmath,amsthm}
\usepackage{indentfirst}
\usepackage{xcolor} 
\usepackage{algorithm}
\usepackage{algpseudocode}

\usepackage{amssymb}
\usepackage{amssymb}
\usepackage{lineno,hyperref}
\usepackage{url}
\usepackage{bm}
\usepackage{overpic}
\usepackage{graphicx}
\usepackage{mathrsfs}
\usepackage{amsfonts}
\usepackage{amssymb}
\usepackage{amsmath,amsthm}
\usepackage{indentfirst}
\usepackage{footmisc}
\usepackage{multirow}
\usepackage{xcolor} 
\usepackage{tikz} 
\usepackage{bm}
\usepackage{algorithm}
\usepackage{algpseudocode}
\usepackage{amsmath}
\usepackage{colortbl}

\usetikzlibrary{arrows,shapes,chains}
\usepackage{pgf}
\usepackage{threeparttable} 
\usepackage{diagbox}
\usetikzlibrary{arrows, decorations.pathmorphing, backgrounds, positioning, fit, petri, automata}
\usepackage{tikz}
\usepackage{tabu} 
\usepackage{float}
\usetikzlibrary{shadows,arrows,positioning,shapes.geometric} 
\pgfdeclarelayer{background}
\pgfdeclarelayer{foreground}
\pgfsetlayers{background,main,foreground}

\begin{document}
\begin{frontmatter}

\title{Application of supervised learning models in the Chinese futures market }
\tnotetext[mytitlenote]{The work is supported by National Natural Science Foundations of China under Grant
11531001.}
\author{Fuquan Tang}
\address{\quad School of Mathematics Science, Shanghai Jiao Tong University, 800 Dongchuan RD. Minhang District, Shanghai, China}

\renewcommand{\thefootnote}{\fnsymbol{footnote}}
\footnotetext{CONTACT Fuquan Tang, Email: dongfang$\_$tang@163.com}

\begin{abstract}
Based on the characteristics of the Chinese futures market, this paper builds a supervised learning model to predict the trend of futures prices and then designs a trading strategy based on the prediction results. The Precision, Recall and F1-score of the classification problem show that our model can meet the accuracy requirements for the classification of futures price movements in terms of test data. The backtest results show that our trading system has an upward trending return curve with low capital retracement.
\end{abstract}


\end{frontmatter}
\section{Introduction}\label{intro}

Renaissance Technologies, Tiger Global Management and many other investment firms have relied on quantitative techniques to achieve success, but their success has not relied solely on econometric models\cite{de2016invited}. Global financial systems have seen considerable growth in size, concentration, and complexity over the past few decades, the complexity of financial systems exceeds the modelling capabilities of traditional quantitative methods. In addition, some very useful data sets, such as satellite images, voice recordings or news sentiment, are beyond the reach of econometrics\cite{lopez2019beyond}. In recent years, many hedge funds have started experimenting with machine learning (ML) methods. ML is a subset of artificial intelligence, where machines are used to learn from previous experience\cite{dunjko2018machine}. Unlike traditional programming, where developers need to predict every potential condition to program, ML's solution can effectively tailor the output to the data. ML algorithm doesn't actually write code, but it builds a computer model of the real world and then trains the model with data. The types of machine learning are: supervised learning, unsupervised learning and reinforcement learning. Supervised learning is the method by which we need to know the correct answer from past data and use it to predict future outcomes. For example, past house prices are used to predict current and future prices. Effectively using a trial-and-error based statistical improvement process, the machine relies on the results of testing against a set of values provided by a supervisor to progressively improve accuracy\cite{ng2015machine,zhang2022longitudinal}. Unsupervised learning is something for which there is no clear-cut right answer here, but we want to make new discoveries from the data. It is most commonly used to sort or group data. For example, the music software Spotify suggests songs or albums you might want to listen to by categorising the music. They then categorise listeners to see if they would prefer to listen to Radiohead or Justin Bieber\cite{lin2018automated,schedl2022music}. Reinforcement learning is something that does not require a domain expert, but requires constant progress towards a predetermined goal. We all know that AphaGo, based on reinforcement learning theory, played a million matches against itself in DeepMind and eventually became the world champion\cite{danesi2020experimental,kim2022clarifying}.

ML has proven to be beneficial in enhancing the services and operations of the financial industry due to the large amount of data that needs to be processed every minute. It is difficult to envisage the modern financial industry not using ML anymore. Kou et al.\cite{kou2019machine}surveyed existing researches and methodologies on assessment and measurement of financial systemic risk combined with machine learning technologies, including big data analysis, network analysis
and sentiment analysis, etc.  Babenko et al.\cite{babenko2021classical} reviewed the classical methods of machine learning (supervised and unsupervised learning), gives examples of the application of different methods and discusses approaches that will be useful for empirical economics research (on data from Ukrainian firms, banks and official state statistics). L{\'o}pez\cite{lopez2019ten} reviewed ten important investment problems for which ML players have delivered superior solutions. These investment issues are: Asset pricing, Risk management and fortfolio contruction, Outliers detection, Bet sizing, Sentiment analysis, Feature importance, Credit ratings and analyst recommendations, Execution, Big data analysis, Controlling for effects and interations. At the same time, ML is not a panacea. while ML technology has its flexibility and effectiveness, it also has a dark side. When misused, ML algorithms can confuse statistical flukes with patterns. Because of the low signal-to-noise ratio that characterises data in finance, it almost ensures that careless users will produce false discoveries at an ever-increasing rate. The rate of failure is particularly high in ML, Prado and L{\'o}pez\cite{lopez20177} summarized the reasons boil down to 7 common errors: The Sisyphus paradigm, Integer differentiation, Inefficient sampling, Wrong labeling, Weighting of non-IID samples, Cross-validation leakage, Backtest overfitting.

There is a vast literature on the attempts of using ML models to predict future trends of financial asset prices. Chiong et al.\cite{chiong2018sentiment} proposed a sentiment analysis-based support vector machine for financial market prediction. Henrique et al.\cite{henrique2019literature} presented a cluster based classifcation model for fnancial crisis prediction. Wu et al.\cite{wu2020labeling} used a new labeling method called ``continuous trend labeling” to label financial time series data and again incorporate price trend. Fleischer\cite{fleischer2019automated} replicated some of Professor Marcos  L{\'o}pez de Prado techniques for the United States Stock Market, including the triple barrier method. However, he did not make use of fractional differentiation. Instead, he utilized integer differentiation and sampled bars around the filings of new 10-K or 10-Q reports, in an attempt of sampling with the arrival of new information. Borges and Rui\cite{2020Ensemble} mentioned Professor Marcos L{\'o}pez de Prado techniques as state of the art, but he does not end up using them, opting for a more straightforward cumulative sum threshold in minute price variation for bar assembling. Galvis\cite{galvis2021data} utilized Professor Marcos L{\'o}pez de Prado data analysis techniques to build a deep neural network for Bitcoin's price direction predictions.

To reduce the common errors of ML models\cite{lopez20177}, this paper replicated some of Professor Marcos  L{\'o}pez de Prado (Cornell University) data analysis techniques for the China Futures Market. Firstly, we use a data transformation method named fractional differentiation to ensure the stationarity of the data while preserving as much memory as possible; Secondly, we use the Triple-Barrier method based on path dependencies to label financial data; Lastly, we build a neural network model  to predict the price movement of the main contract in the Chinese futures market. To have a more tangible result, a trade simulation was executed on the test data.

\section{Data Analysis}\label{data}
The main application of ML in the financial sector is to use data from market transactions, such as prices and trading volumes, to predict the movement of asset prices. According to the literature\cite{dhifaoui2022determinism}, stock price movements are non-linear and stochastic. In particular, trading activity is rarely consistent over the course of a day, week or month. Market trading behaviour can vary with the flow of information in the form of macroeconomic data releases, industry policy documents or company specific announcements. The following Figure \ref{fig-1}$\verb|--|$\ref{fig-2} show the autocorrelation (ACF) and partial autocorrelation (PACF) plots of closing price data for the main rebar contract on the China Futures Market, the ACF and PACF plots indicate that the time series is non-stationary.
\begin{figure}
\begin{minipage}[t]{0.45\linewidth}
\centering
\includegraphics[width=\textwidth]{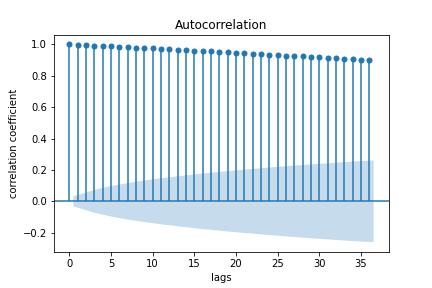}
\caption{The autocorrelation of rebar contract} 
\label{fig-1}
\end{minipage}
\begin{minipage}[t]{0.45\linewidth}
\centering
\includegraphics[width=\textwidth]{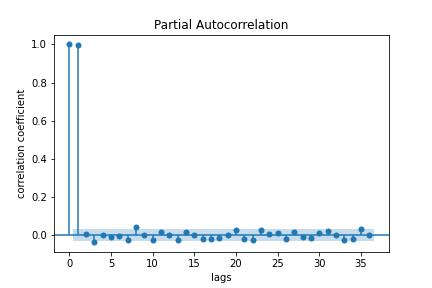}
\caption{The partial autocorrelation of rebar contract}
\label{fig-2}
\end{minipage}
\end{figure} 

Our objective is to build supervised learning models to predict the price movements of financial time series. In the data analysis part, we do two main tasks: firstly, we convert the time series into an approximately smooth time series by fractional difference transformation; secondly, we use the Triple-Barrer Method to label the sample data.

\subsection{Fractionally Differentiated Features}
It is known that financial series exhibit low signal-to-noise ratios\cite{de2015future} and financial data are unstable\cite{gadanecz2008measures,mamedio2019strategic}. For data analysis and ML, they have to be transformed into stationary. Our usual practice is to first take the logarithm of the time series data and then differentiate it by integers, but this is problematic? In order to ensure a stationary time series, the memory of the time series is discarded, as Marcos  L{\'o}pez de Prado\cite{de2018advances}stated that there is memory between prices, today's prices are based on previous prices. In this paper, we apply Fractionally Differentiated Features\cite{de2018advances} to perform a stationary transformation on the price series of futures.

Consider the backshift operator, $B$, applied to a matrix of real-valued features $\left\{X_{t}\right\}$, where $B^{k} X_{t}=X_{t-k}$ for any integer $k \geq 0$. For example, $(1-B)^{2}=1-2 B+B^{2}$, where $B^{2} X_{t}=X_{t-2}$, so that $(1-B)^{2} X_{t}=X_{t}-2 X_{t-1}+X_{t-2}$. 

Note that $(x+y)^{n}=$ $\sum_{k=0}^{n}\left(\begin{array}{l}n \\ k\end{array}\right) x^{k} y^{n-k}=\sum_{k=0}^{n}\left(\begin{array}{l}n \\ k\end{array}\right) x^{n-k} y^{k}$, where  $n$ is a positive integer. For a real number $d,(1+x)^{d}=\sum_{k=0}^{\infty}\left(\begin{array}{l}d \\ k\end{array}\right) x^{k}$ is a binomial series. In a fractional model, the exponent d is allowed to be a real number, with the
following formal binomial series expansion:
\begin{equation}
\begin{aligned}
(1-B)^{d}=\sum_{k=0}^{\infty}\left(\begin{array}{l}
d \\
k
\end{array}\right)(-B)^{k} &=\sum_{k=0}^{\infty} \frac{\prod_{i=0}^{k-1}(d-i)}{k !}(-B)^{k} \\
&=\sum_{k=0}^{\infty}(-B)^{k} \prod_{i=0}^{k-1} \frac{d-i}{k-i} \\
&=1-d B+\frac{d(d-1)}{2 !} B^{2}-\frac{d(d-1)(d-2)}{3 !} B^{3}+\cdots.
\end{aligned}
\end{equation}

In the following we analyse how memory is preserved for d-order fractional order differences. This arithmetic series consists of a dot product
series consists of a dot product
\begin{equation}
\tilde{X}_{t}=\sum_{k=0}^{\infty} \omega_{k} X_{t-k}
\end{equation}
with weights $\omega$
\begin{equation}
\omega=\left\{1,-d, \frac{d(d-1)}{2 !},-\frac{d(d-1)(d-2)}{3 !}, \ldots,(-1)^{k} \prod_{i=0}^{k-1} \frac{d-i}{k !}, \ldots\right\}
\end{equation}
and values $X$
\begin{equation}
X=\left\{X_{t}, X_{t-1}, X_{t-2}, X_{t-3}, \ldots, X_{t-k}, \ldots\right\}
\end{equation}
When $d$ is a positive integer number, $\prod_{i=0}^{k-1} \frac{d-i}{k !}=0, \forall k>d$, and memory beyond that point is cancelled. For example, $d=1$ is used to compute returns, where $\prod_{i=0}^{k-1} \frac{d-i}{k !}=0, \forall k>1$, and $\omega=\{1,-1,0,0, \ldots\}$.
Looking at the sequence of weights, $\omega$, we can appreciate that for $k = 0,\cdots,\infty$, with
$\omega_0= 1$, the weights can be generated iteratively as:
\begin{equation}
\omega_{k}=-\omega_{k-1} \frac{d-k+1}{k}
\end{equation}
Let us consider the convergence of the weights. From the above result, we can see
that for $k>d$, if $\omega_{k-1}\neq 0$, then $\left|\frac{\omega_{k}}{\omega_{k-1}}\right|=\left|\frac{d-k+1}{k}\right|<1$, and $\omega_{k}=0$, otherwise. Consequently, the weights converge asymptotically to zero, as an infinite product of factors within the unit circle. Also, for a positive $d$ and $k < d + 1$, we have $\frac{d-k+1}{k}\geq 0$, which makes the initial weights alternate in sign. For a non-integer $d$, once $k \geq d + 1$, $\omega_{k}$, will be negative if int $[d]$ is even, and positive otherwise. Summarizing, $\lim _{k \rightarrow \infty} \omega_{k}=0^{-}$ (converges to zero from the left) when int $[d]$ is even, and $\lim _{k \rightarrow \infty} \omega_{k}=0^{+}$ (converges to zero from the right) when Int $[d]$ is odd. In the special case $d\in(0, 1)$, this means that $-1< \omega_{k}<0,\forall k>0$. This alternation of weight signs is necessary to make $\{\tilde{X}_{t}\}_{t=1, \ldots, T}$ stationary, as memory wanes or is offset over the long run. There are two alternative implementations of fractional differentiation: the standard ``expanding window” method, and a new method that I call ``fixed-width window fracdiff” (FFD)\cite{de2018advances}.

In the framework of our algorithm, we apply the Fixed-Width Window Fracdiff method. We drop the weights after their modulus $(|\omega_{k}|)$ falls below a given threshold value $(\tau)$, this is equivalent to find the first $l^{*}$ such that $|\omega_{l^{*}}|\geq\tau$ and $|\omega_{l^{*}+1}|\leq\tau$. Then we obtain a new variable $\tilde{\omega}_k$
\begin{equation}
\tilde{\omega}_{k}= \begin{cases}\omega_{k} & \text { if } k \leq l^{*} \\ 0 & \text { if } k>l^{*}\end{cases}
\end{equation}
and $\tilde{X}_{t}=\sum_{k=0}^{t^{*}} \tilde{\omega}_{k} X_{t-k}$, for $t=T-l^{*}+1, \ldots, T$.
This procedure has the advantage that the same vector of weights is used
across all estimates of $\tilde{X}_{t},t=l^{*},\cdots, T$, hence avoiding the negative drift caused by an expanding window’s added weights. The result is a driftless blend of level plus noise, as expected. The distribution is no longer Gaussian, as a result of the skewness and excess kurtosis that comes with memory, however it is stationary.

\begin{figure}
\begin{minipage}[t]{0.46\linewidth}
\centering
\includegraphics[width=\textwidth]{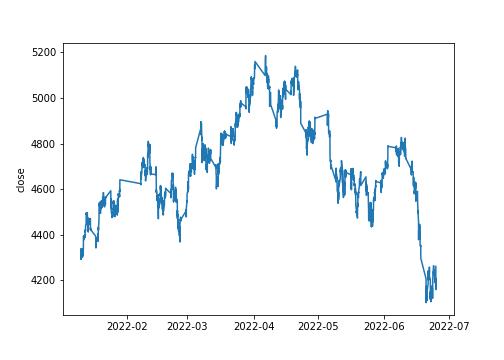}
\caption{The closing price curves of rebar contract} 
\label{fig-3}
\end{minipage}
\begin{minipage}[t]{0.50\linewidth}
\centering
\includegraphics[width=\textwidth]{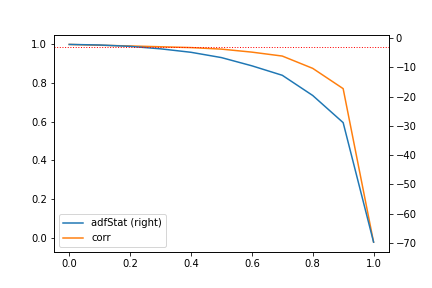}
\caption{The ADF statistic curves of rebar contract}
\label{fig-4}
\end{minipage}
\end{figure} 
Figure \ref{fig-3} is the closeing price curves of Rebar contract and  Figure \ref{fig-4} is the ADF statistic curves for Rebar futures based on different differential orders. In Figure \ref{fig-4}, On the right y-axis, it plots the Augmented Dickey-Fuller (ADF) statistic computed on Rebar futures log-prices. On the x-axis, it displays the $d$ value used to generate the series on which the ADF statistic was computed. The original series has an ADF statistic of -0.3387, while the returns series has an ADF statistic of -46.9114. At a $95\%$ confidence level, the test's critical value is -2.8618. The ADF statistic crosses that threshold in the vicinity of $d = 0.3$. The left y-axis plots the correlation between the original series
$(d = 0)$ and the differentiated series at various d values. At $d=0.3$, the correlation coefficient of the time series is 0.988. This confirms that the FFD has been successful in achieving stationarity without giving up too much memory. In contrast, the correlation between the original series and the returns series is only 0.022, hence  the standard integer differentiation wipes out the series' memory almost entirely. Virtually all finance papers attempt to recover stationarity by applying an integer differentiation $d = 1 \gg0.3$, which means that most studies have over-differentiated the series, they have removed much more memory than was necessary to satisfy standard econometric assumptions.
\begin{figure}
\begin{minipage}[t]{0.45\linewidth}
\centering
\includegraphics[width=\textwidth]{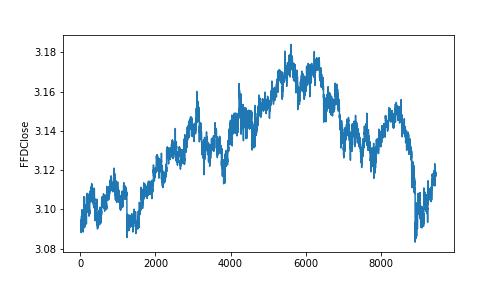}
\caption{The FFD of the rebar contract with d=0.3} 
\label{fig-5}
\end{minipage}
\begin{minipage}[t]{0.45\linewidth}
\centering
\includegraphics[width=\textwidth]{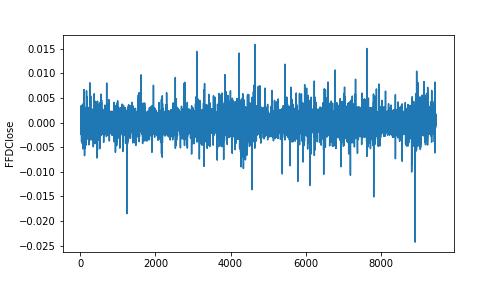}
\caption{The FFD of the rebar contract with d=1}
\label{fig-6}
\end{minipage}
\end{figure} 

 Figure \ref{fig-5}$\verb|--|$\ref{fig-6} are the FFD of the main rebar contract  with $d=0.3$ and $d=1$. The FFD with $d=0.3$ is more similar to the original time series in Figure \ref{fig-3}, and the FFD with $d=1$ resembles white noise and differs more from the original time series.

\begin{table}
  \centering
 \setlength{\tabcolsep}{3mm}{%
\resizebox{\linewidth}{!}{
\scalebox{0.06}{
\renewcommand{\arraystretch}{1.2}
    \begin{tabular}{|c |c c c c c c c c c c c|}
    \hline
   & 0 & 0.1& 0.2& 0.3& 0.4& 0.5& 0.6& 0.7 & 0.8& 0.9&1\cr\cline{1-12}

  Iron ores & -0.374 & -1.651& -2.246& \cellcolor{yellow}-3.139&  \cellcolor{yellow}-4.309&  \cellcolor{yellow}-6.040& \cellcolor{yellow} -8.691& \cellcolor{yellow} -11.774 & \cellcolor{yellow} -18.417&  \cellcolor{yellow}-27.614& \cellcolor{yellow}-78.621\cr
  Rebar & -2.079 & -2.403& \cellcolor{yellow}-3.049&  \cellcolor{yellow}-4.156&  \cellcolor{yellow}-5.632& \cellcolor{yellow} -7.860& \cellcolor{yellow} -11.267&  \cellcolor{yellow}-15.182 &  \cellcolor{yellow}-23.465&  \cellcolor{yellow}-34.486& \cellcolor{yellow}-78.921\cr
Hot rolled coils  & -1.690 & -2.060& -2.771&\cellcolor{yellow} -3.889& \cellcolor{yellow} -5.361& \cellcolor{yellow} -7.540&  \cellcolor{yellow}-10.852& \cellcolor{yellow} -14.656 &  \cellcolor{yellow}-22.709&  \cellcolor{yellow}-33.473& \cellcolor{yellow}-78.978\cr
Stainless steel  & -1.830 & -2.071& -2.638&\cellcolor{yellow} -3.495&  \cellcolor{yellow}-4.699& \cellcolor{yellow} -6.501&  \cellcolor{yellow}-9.281& \cellcolor{yellow} -12.537 &  \cellcolor{yellow}-19.680&  \cellcolor{yellow}-29.717& \cellcolor{yellow}-93.928\cr
Ferro Silicon & -1.966 & -2.168& -2.672& \cellcolor{yellow}-3.488& \cellcolor{yellow}-4.623& \cellcolor{yellow}-6.331& \cellcolor{yellow}-8.979& \cellcolor{yellow}-12.024 &\cellcolor{yellow} -18.528& \cellcolor{yellow}-27.166&\cellcolor{yellow}-62.703\cr
Manganese silicon & -2.240 & -2.558& \cellcolor{yellow}-3.257& \cellcolor{yellow}-4.305& \cellcolor{yellow}-5.767& \cellcolor{yellow}-7.911& \cellcolor{yellow}-11.203& \cellcolor{yellow}-14.944 & \cellcolor{yellow}-22.643& \cellcolor{yellow}-32.366&\cellcolor{yellow}-62.569\cr
Shanghai Copper & -2.354 & -2.592&\cellcolor{yellow} -3.158& \cellcolor{yellow}-4.099&\cellcolor{yellow} -5.417& \cellcolor{yellow}-7.428& \cellcolor{yellow}-10.519& \cellcolor{yellow}-14.148 &\cellcolor{yellow} -21.988& \cellcolor{yellow}-32.780&\cellcolor{yellow}-89.515\cr
Shanghai Aluminium & -1.744 & -1.939& -2.368& \cellcolor{yellow}-3.073& \cellcolor{yellow}-4.070& \cellcolor{yellow}-5.584&\cellcolor{yellow} -7.928& \cellcolor{yellow}-10.714 & \cellcolor{yellow}-16.764& \cellcolor{yellow}-25.439&\cellcolor{yellow}-90.908\cr
Shanghai Zinc & -1.906 & -2.233& \cellcolor{yellow}-2.919&\cellcolor{yellow} -3.962& \cellcolor{yellow}-5.374& \cellcolor{yellow}-7.461& \cellcolor{yellow}-10.657& \cellcolor{yellow}-14.355 & \cellcolor{yellow}-22.326& \cellcolor{yellow}-33.259&\cellcolor{yellow}-89.992\cr
Shanghai Lead &\cellcolor{yellow} -4.584 &\cellcolor{yellow} -5.550& \cellcolor{yellow}-7.381&\cellcolor{yellow} -10.119&\cellcolor{yellow} -13.667&\cellcolor{yellow} -18.750&\cellcolor{yellow} -26.154& \cellcolor{yellow}-34.291 & \cellcolor{yellow}-48.889&\cellcolor{yellow} -64.613&\cellcolor{yellow}-92.167\cr
Shanghai Nickel & -1.487 & -1.759& -2.327&\cellcolor{yellow} -3.196& \cellcolor{yellow}-4.363& \cellcolor{yellow}-6.101&\cellcolor{yellow} -8.768&\cellcolor{yellow} -11.897 &\cellcolor{yellow} -18.720& \cellcolor{yellow}-28.326&\cellcolor{yellow}-93.281\cr
Shanghai Tin & -1.564 & -1.586& -1.722& -2.055& -2.572&\cellcolor{yellow} -3.415& \cellcolor{yellow}-4.769& \cellcolor{yellow}-6.387 &\cellcolor{yellow} -10.017& \cellcolor{yellow}-15.297&\cellcolor{yellow}-86.878\cr
International Copper & -2.266 & -2.505& \cellcolor{yellow}-3.070& \cellcolor{yellow}-3.995& \cellcolor{yellow}-5.297& \cellcolor{yellow}-7.274&\cellcolor{yellow} -10.315& \cellcolor{yellow}-13.888 & \cellcolor{yellow}-21.606& \cellcolor{yellow}-32.258&\cellcolor{yellow}-89.528\cr
Shanghai Gold & -2.458 & \cellcolor{yellow}-2.966& \cellcolor{yellow}-4.002&\cellcolor{yellow} -5.573&\cellcolor{yellow} -7.638&\cellcolor{yellow} -10.680&\cellcolor{yellow} -15.31& \cellcolor{yellow}-20.631 & \cellcolor{yellow}-31.752& \cellcolor{yellow}-46.377&\cellcolor{yellow}-101.248\cr
Bean One &\cellcolor{yellow} -3.791 &\cellcolor{yellow} -4.311&\cellcolor{yellow} -5.351&\cellcolor{yellow} -7.042& \cellcolor{yellow}-9.362& \cellcolor{yellow}-12.810& \cellcolor{yellow}-18.013&\cellcolor{yellow} -23.870 & \cellcolor{yellow}-35.332&\cellcolor{yellow} -48.882&\cellcolor{yellow}-80.202\cr
Bean Two & -2.107 & -2.314& -2.852& \cellcolor{yellow}-3.708&\cellcolor{yellow} -4.895& \cellcolor{yellow}-6.694& \cellcolor{yellow}-9.497& \cellcolor{yellow}-12.740 &\cellcolor{yellow} -19.811&\cellcolor{yellow} -29.479&\cellcolor{yellow}-77.838\cr
Soybean oil & -1.827 & -2.019& -2.511&\cellcolor{yellow} -3.279& \cellcolor{yellow}-4.351&\cellcolor{yellow} -5.986& \cellcolor{yellow}-8.480&\cellcolor{yellow} -11.395 & \cellcolor{yellow}-17.791& \cellcolor{yellow}-26.665&\cellcolor{yellow}-77.745\cr
Soybean meal & -1.937 & -2.203& -2.800&\cellcolor{yellow} -3.760&\cellcolor{yellow} -5.017& \cellcolor{yellow}-6.932& \cellcolor{yellow}-9.885& \cellcolor{yellow}-13.310 & \cellcolor{yellow}-20.677& \cellcolor{yellow}-30.741&\cellcolor{yellow}-79.534\cr
Vegetable oil & -1.858 & -2.030& -2.430& \cellcolor{yellow}-3.118&\cellcolor{yellow} -4.092&\cellcolor{yellow} -5.589& \cellcolor{yellow}-7.922& \cellcolor{yellow}-10.653 & \cellcolor{yellow}-16.647&\cellcolor{yellow} -25.071&\cellcolor{yellow}-78.483\cr
Vegetable meal & -2.092 & -2.345& \cellcolor{yellow}-2.897& \cellcolor{yellow}-3.826&\cellcolor{yellow} -5.059& \cellcolor{yellow}-6.958&\cellcolor{yellow} -9.893& \cellcolor{yellow}-13.292 & \cellcolor{yellow}-20.630& \cellcolor{yellow}-30.646&\cellcolor{yellow}-78.871\cr
Palm oil & -1.623 & -1.771& -2.155& -2.791& \cellcolor{yellow}-3.685&\cellcolor{yellow} -5.056&\cellcolor{yellow} -7.173&\cellcolor{yellow} -9.644 &\cellcolor{yellow} -15.128& \cellcolor{yellow}-22.870&\cellcolor{yellow}-77.838\cr
Corn & -1.765 & -2.046& -2.640&\cellcolor{yellow} -3.570&\cellcolor{yellow} -4.824& \cellcolor{yellow}-6.695& \cellcolor{yellow}-9.579&\cellcolor{yellow} -12.925 & \cellcolor{yellow}-20.128&\cellcolor{yellow} -30.000&\cellcolor{yellow}-80.550\cr
Corn Starch & -2.273 & -2.572&\cellcolor{yellow} -3.234&\cellcolor{yellow} -4.312& \cellcolor{yellow}-5.765& \cellcolor{yellow}-7.969&\cellcolor{yellow} -11.349&\cellcolor{yellow} -15.259 &\cellcolor{yellow} -23.566& \cellcolor{yellow}-34.661&\cellcolor{yellow}-80.752\cr
Eggs & -2.652 & \cellcolor{yellow}-2.947&\cellcolor{yellow} -3.682& \cellcolor{yellow}-4.894& \cellcolor{yellow}-6.521& \cellcolor{yellow}-8.947& \cellcolor{yellow}-12.566&\cellcolor{yellow} -16.684 & \cellcolor{yellow}-24.942& \cellcolor{yellow}-34.955&\cellcolor{yellow}-61.599\cr
No.1 Cotton & -1.459 & -1.544& -1.820& -2.295&\cellcolor{yellow} -2.991&\cellcolor{yellow} -4.050& \cellcolor{yellow}-5.728&\cellcolor{yellow} -7.712 &\cellcolor{yellow} -12.121&\cellcolor{yellow} -18.468&\cellcolor{yellow}-77.492\cr
Cotton Yarn & -1.813 & -1.913& -2.278&\cellcolor{yellow} -2.876&\cellcolor{yellow} -3.755& \cellcolor{yellow}-5.084& \cellcolor{yellow}-7.199& \cellcolor{yellow}-9.701 & \cellcolor{yellow}-15.203& \cellcolor{yellow}-23.017&\cellcolor{yellow}-77.862\cr
White sugar & -2.850 & \cellcolor{yellow}-3.024& \cellcolor{yellow}-3.550& \cellcolor{yellow}-4.526&\cellcolor{yellow} -5.885& \cellcolor{yellow}-8.001&\cellcolor{yellow} -11.299&\cellcolor{yellow} -15.096 & \cellcolor{yellow}-23.256&\cellcolor{yellow} -34.188&\cellcolor{yellow}-79.838\cr
Apple & -0.491 & -0.733& -1.177& -1.803& -2.579&\cellcolor{yellow} -3.697& \cellcolor{yellow}-5.378& \cellcolor{yellow}-7.307 & \cellcolor{yellow}-11.504& \cellcolor{yellow}-17.424&\cellcolor{yellow}-61.124\cr
Red dates & -1.229 & -1.393& -1.758& -2.353&\cellcolor{yellow} -3.151& \cellcolor{yellow}-4.355&\cellcolor{yellow} -6.222& \cellcolor{yellow}-8.409 &\cellcolor{yellow} -13.178&\cellcolor{yellow} -19.858&\cellcolor{yellow}-62.742\cr
Peanuts & -1.395 & -1.646& -2.128&\cellcolor{yellow} -2.870& \cellcolor{yellow}-3.859&\cellcolor{yellow} -5.344& \cellcolor{yellow}-7.615&\cellcolor{yellow} -10.254 &\cellcolor{yellow} -15.930& \cellcolor{yellow}-23.642&\cellcolor{yellow}-60.799\cr
Stalked rice & -0.942 & -1.222& -1.775& -2.597& \cellcolor{yellow}-3.632&\cellcolor{yellow} -5.161& \cellcolor{yellow}-7.499& \cellcolor{yellow}-10.210 &\cellcolor{yellow} -16.114& \cellcolor{yellow}-24.446&\cellcolor{yellow}-81.155\cr
Pig & -1.966 & -1.760& -1.702& -1.784& -2.242& \cellcolor{yellow}-2.868& \cellcolor{yellow}-3.975& \cellcolor{yellow}-5.281 &\cellcolor{yellow} -8.243&\cellcolor{yellow} -12.613&\cellcolor{yellow}-60.838\cr
Fuel & -1.428 & -1.615& -2.019& -2.704& \cellcolor{yellow}-3.618& \cellcolor{yellow}-5.016& \cellcolor{yellow}-7.184&\cellcolor{yellow} -9.724 &\cellcolor{yellow} -15.279& \cellcolor{yellow}-23.158&\cellcolor{yellow}-79.110\cr
Liquefied Petroleum Gas & -1.353 & -1.570& -2.036& -2.769&\cellcolor{yellow} -3.750& \cellcolor{yellow}-5.213& \cellcolor{yellow}-7.478& \cellcolor{yellow}-10.109 & \cellcolor{yellow}-15.865& \cellcolor{yellow}-23.974&\cellcolor{yellow}-78.700\cr
Ethylene glycol & -2.809 &\cellcolor{yellow} -3.234&\cellcolor{yellow} -4.156&\cellcolor{yellow} -5.650& \cellcolor{yellow}-7.603&\cellcolor{yellow} -10.516& \cellcolor{yellow}-14.920& \cellcolor{yellow}-19.927 & \cellcolor{yellow}-30.077&\cellcolor{yellow} -42.728&\cellcolor{yellow}-79.186\cr
Polyethylene & -2.615 &\cellcolor{yellow} -3.069& \cellcolor{yellow}-4.063&\cellcolor{yellow} -5.545& \cellcolor{yellow}-7.515& \cellcolor{yellow}-10.414&\cellcolor{yellow} -14.779& \cellcolor{yellow}-19.733 & \cellcolor{yellow}-29.808&\cellcolor{yellow} -42.343&\cellcolor{yellow}-78.727\cr
PTA & -1.728 & -1.870& -2.268&\cellcolor{yellow} -2.961& \cellcolor{yellow}-3.392& \cellcolor{yellow}-5.391& \cellcolor{yellow}-7.680&\cellcolor{yellow} -10.369 & \cellcolor{yellow}-16.236&\cellcolor{yellow} -24.498&\cellcolor{yellow}-79.117\cr
Polyvinyl chloride & -1.502 & -1.905& -2.792& \cellcolor{yellow}-4.053&\cellcolor{yellow} -5.670& \cellcolor{yellow}-8.015& \cellcolor{yellow}-11.542&\cellcolor{yellow} -15.581 & \cellcolor{yellow}-24.040& \cellcolor{yellow}-35.189&\cellcolor{yellow}-79.259\cr
Polypropylene &\cellcolor{yellow} -3.028 & \cellcolor{yellow}-3.635& \cellcolor{yellow}-4.872& \cellcolor{yellow}-6.724& \cellcolor{yellow}-9.145&\cellcolor{yellow} -12.659& \cellcolor{yellow}-17.878& \cellcolor{yellow}-23.733 &\cellcolor{yellow} -35.080& \cellcolor{yellow}-48.368&\cellcolor{yellow}-78.921\cr
Styrene & -2.438 & -2.555& \cellcolor{yellow}-3.124&\cellcolor{yellow} -4.137&\cellcolor{yellow} -5.491&\cellcolor{yellow} -7.554& \cellcolor{yellow}-10.763& \cellcolor{yellow}-14.514 & \cellcolor{yellow}-22.418& \cellcolor{yellow}-33.132&\cellcolor{yellow}-79.563\cr
Pure soda & -2.377 & -2.384& -2.543&\cellcolor{yellow} -2.966& \cellcolor{yellow}-3.692& \cellcolor{yellow}-4.866& \cellcolor{yellow}-6.771& \cellcolor{yellow}-9.041 & \cellcolor{yellow}-14.088& \cellcolor{yellow}-21.360&\cellcolor{yellow}-77.330\cr
Urea & -2.068 & -2.300& -2.748&\cellcolor{yellow} -3.541& \cellcolor{yellow}-4.648&\cellcolor{yellow} -6.345& \cellcolor{yellow}-8.960& \cellcolor{yellow}-11.991 & \cellcolor{yellow}-18.450& \cellcolor{yellow}-27.080&\cellcolor{yellow}-62.846\cr
Rubber & -2.720 & \cellcolor{yellow}-3.325& \cellcolor{yellow}-4.504& \cellcolor{yellow}-6.262& \cellcolor{yellow}-8.566& \cellcolor{yellow}-11.913&\cellcolor{yellow} -16.925& \cellcolor{yellow}-22.584 & \cellcolor{yellow}-33.727& \cellcolor{yellow}-47.110&\cellcolor{yellow}-80.432\cr
No.20 rubber &\cellcolor{yellow} -4.229 & \cellcolor{yellow}-4.997& \cellcolor{yellow}-6.617& \cellcolor{yellow}-9.088& \cellcolor{yellow}-12.283& \cellcolor{yellow}-16.876& \cellcolor{yellow}-22.508& \cellcolor{yellow}-30.773 &\cellcolor{yellow} -43.653& \cellcolor{yellow}-57.311&\cellcolor{yellow}-80.153\cr
Paper pulp & -1.998 & -2.178& -2.657&\cellcolor{yellow} -3.442& \cellcolor{yellow}-4.548& \cellcolor{yellow}-6.236& \cellcolor{yellow}-8.869& \cellcolor{yellow}-11.921 & \cellcolor{yellow}-18.590& \cellcolor{yellow}-27.822&\cellcolor{yellow}-78.333\cr
Staple Fiber & -2.565 &\cellcolor{yellow} -2.876& \cellcolor{yellow}-3.644& \cellcolor{yellow}-4.885& \cellcolor{yellow}-6.526& \cellcolor{yellow}-8.997&\cellcolor{yellow} -12.783& \cellcolor{yellow}-17.152 &\cellcolor{yellow} -26.248& \cellcolor{yellow}-38.022&\cellcolor{yellow}-78.730\cr
Coking coal & -1.676 & -1.840& -2.240& \cellcolor{yellow}-2.919&\cellcolor{yellow} -3.860&\cellcolor{yellow} -5.297& \cellcolor{yellow}-7.543& \cellcolor{yellow}-10.172 & \cellcolor{yellow}-15.957& \cellcolor{yellow}-24.110&\cellcolor{yellow}-78.567\cr
Coke & -1.866 & -2.188& -2.858& \cellcolor{yellow}-3.906& \cellcolor{yellow}-5.304&\cellcolor{yellow} -7.382& \cellcolor{yellow}-10.562&\cellcolor{yellow} -14.237 &\cellcolor{yellow} -22.057& \cellcolor{yellow}-32.601&\cellcolor{yellow}-79.257\cr
CSI & -1.706 & -2.020& -2.642&\cellcolor{yellow} -3.608& \cellcolor{yellow}-4.899&\cellcolor{yellow} -6.806& \cellcolor{yellow}-9.714&\cellcolor{yellow} -13.073 & \cellcolor{yellow}-20.109&\cellcolor{yellow} -29.431&\cellcolor{yellow}-66.374\cr
Shanghai and Shenzhen & -1.142 & -1.420& -1.971& -2.791&\cellcolor{yellow} -3.854& \cellcolor{yellow}-5.415& \cellcolor{yellow}-7.795& \cellcolor{yellow}-10.552 &\cellcolor{yellow} -16.467&\cellcolor{yellow} -24.594&\cellcolor{yellow}-67.359\cr
SSE & -1.239 & -1.511& -2.051& \cellcolor{yellow}-2.865&\cellcolor{yellow} -3.931&\cellcolor{yellow} -5.503& \cellcolor{yellow}-7.908&\cellcolor{yellow} -10.693 &\cellcolor{yellow} -16.678& \cellcolor{yellow}-24.879&\cellcolor{yellow}-67.436\cr
Ten debts & -1.654 & -1.836& -2.302&\cellcolor{yellow} -3.054&\cellcolor{yellow} -4.101&\cellcolor{yellow} -5.657& \cellcolor{yellow}-8.075&\cellcolor{yellow} -10.890 & \cellcolor{yellow}-16.922&\cellcolor{yellow} -25.262&\cellcolor{yellow}-67.843\cr
Five debts & -1.687 & -1.858& -2.328& \cellcolor{yellow}-3.080&\cellcolor{yellow} -4.141& \cellcolor{yellow}-5.704& \cellcolor{yellow}-8.136&\cellcolor{yellow} -10.976 & \cellcolor{yellow}-17.042& \cellcolor{yellow}-25.401&\cellcolor{yellow}-67.324\cr
Second Debt & -1.550 & -1.769& -2.293&\cellcolor{yellow} -3.127& \cellcolor{yellow}-4.248& \cellcolor{yellow}-5.911& \cellcolor{yellow}-8.479& \cellcolor{yellow}-11.462 & \cellcolor{yellow}-17.814&\cellcolor{yellow} -26.502&\cellcolor{yellow}-68.633\cr
Shanghai Silver & -1.724 & -2.337&\cellcolor{yellow} -3.494& \cellcolor{yellow}-5.141& \cellcolor{yellow}-7.226& \cellcolor{yellow}-10.234& \cellcolor{yellow}-14.766&\cellcolor{yellow} -19.955 &\cellcolor{yellow} -30.795& \cellcolor{yellow}-45.103&\cellcolor{yellow}-101.618\cr
Crude Oil & -1.398 & -1.521& -1.833& -2.389& \cellcolor{yellow}-3.161& \cellcolor{yellow}-4.350&\cellcolor{yellow} -6.203& \cellcolor{yellow}-8.392 & \cellcolor{yellow}-13.252& \cellcolor{yellow}-20.301&\cellcolor{yellow}-100.094\cr
Low sulphur crude & -0.759 & -1.015& -1.490& -2.197& \cellcolor{yellow}-3.083& \cellcolor{yellow}-4.385&\cellcolor{yellow} -6.367& \cellcolor{yellow}-8.658 &\cellcolor{yellow} -13.660& \cellcolor{yellow}-20.799&\cellcolor{yellow}-78.810\cr
Petroleum bitumen & -1.114 & -1.440& -2.068&\cellcolor{yellow} -2.987& \cellcolor{yellow}-4.164& \cellcolor{yellow}-5.885& \cellcolor{yellow}-8.497& \cellcolor{yellow}-11.516 & \cellcolor{yellow}-18.017& \cellcolor{yellow}-27.019&\cellcolor{yellow}-78.242\cr
Methanol & -2.150 & -2.532&\cellcolor{yellow} -3.357& \cellcolor{yellow}-4.626&\cellcolor{yellow} -6.293& \cellcolor{yellow}-8.753& \cellcolor{yellow}-12.482& \cellcolor{yellow}-16.762 & \cellcolor{yellow}-25.667& \cellcolor{yellow}-37.210&\cellcolor{yellow}-78.386\cr
Glass & -0.802 & -1.117& -1.701& -2.545& \cellcolor{yellow}-3.612&\cellcolor{yellow} -5.167&\cellcolor{yellow} -7.524& \cellcolor{yellow}-10.256 &\cellcolor{yellow} -16.172&\cellcolor{yellow} -24.481&\cellcolor{yellow}-79.247\cr\cline{1-12}
    \end{tabular}}
}}
\vskip 0.3cm
\caption{ Optimal FradDiff stationarity of contract indices for china futures trading varieties.}
\label{tab-1}
\end{table}
Table \ref{tab-1} shows the ADF statistics after FFD(d) for 60 major futures contract indices in the Chinese futures market. We know that the ADF value to satisfy the unit root test is -2.8623, the yellow part of the cell in the table indicates that the unit root test can be satisfied. From the table, we can see that $Shanghai\,Lead$, $Bean\,One$, $Polypropylene$ and $No.20\,rubber$ do not need to be differenced to satisfy the unit root test;
Most futures index contracts only need to be differenced with order $d=0.3$ to obtain a smooth series, only $Shanghai\,Tin$, $Apple$, and $Pig$ need to be differenced with order $d=0.5$ to satisfy the unit root test. 
In all cases, the standard $d = 1$ used for computing returns implies over-differentiation.

\subsection{Labeling}
Regarding the labelling of data in financial time series, most of the supervised learning literature uses the fixed time horizon method (FHM) to label observations. This method can be described as follows. Consider a features matrix $X$ with $I$ rows,  $\left\{X_{t}\right\}_{t=1, \ldots, I}$ , drawn from some barswith index $t = 1,\cdots T$, where $I\leq T$. An observation $X_i$ is assigned a label $y_{i}\in\{-1,0,1\}$.

\begin{equation}
y_{i}=\left\{\begin{array}{cl}
-1 & \text { if } r_{t_{i, 0}, t_{i, 0}+h}<-\tau \\
0 & \text { if }\left|r_{t_{i, 0}, t_{i, 0}+h}\right| \leq \tau \\
1 & \text { if } r_{t_{i, 0}, t_{i, 0}+h}>\tau
\end{array}\right.
\end{equation}
where $\tau$ is a pre-defined constant threshold, $t_{i,0}$ is the index of the bar immediately after $X_{i}$ takes place, $t_{i,0}+h$ is the index of the $h-$th bar after $t_{i,0}$, and $r_{t_{i,0},t_{i,0}+h}$ is the price return over a bar horizon $h$,
\begin{equation}
r_{t_{i, 0}, t_{i, 0}+h}=\frac{p_{t_{i, 0}+h}}{p_{t_{i, 0}}}-1
\end{equation}
Because the literature almost always works with time bars, $h$ implies a fixed-time horizon. FHM has two main problems: firstly, $\tau$ is a constant, the volatility of the underlying tends to change over time. When the volatility of the underlying is large, few samples are labelled as zero, while when the volatility of the underlying is small, most samples are labelled as zero. In other words, the samples are labelled solely on the basis of the absolute magnitude of the value rather than being statistically significant, which is clearly unreasonable. Furthermore, FHM does not take into account stop-loss and take-profit issues, but rather a general indication of the overall return over the time interval. For situations where a stop-loss has been triggered or a position has been forcibly closed, the mark-up based on total returns often does not reflect the true situation. Marcos  L{\'o}pez de Prado\cite{de2018advances} proposed an innovative approach to label examples - the Triple-Barrier Method. This is a path dependence approach , and is therefore an effective solution to the stop-loss problem mentioned above, the details of this method are described below.

\subsubsection{The Triple-Barrer Method}
I call it the triple barrier method because it observes whether three barriers are touched based on the labels. First, we set two horizontal barriers and one vertical barrier. The two horizontal barriers are limited by the profit-taking and stop-loss definitions and are dynamic functions of estimated volatility (whether realised or implied), the third barrier is defined by the number of bars that have passed since (the expiry limit). If the upper barrier is touched first, we mark the observation as 1. If the lower barrier is touched first, we mark the observation as -1. If the vertical barrier is touched first, it is marked as 0. Overall, the horizontal label defines which price level will be classified as positive or negative, while the vertical label determines how far back in the label you will look.
\begin{equation}
Y_{i}=\left\{\begin{aligned}
-1 & ~~\text { if lower barrier is first touched }, \\
0 & ~~\text { if no previous barriers were reached }, \\
1 & ~~\text { if upper barrier is first touched },
\end{aligned}\right.
\end{equation}
where upper barrie and lower barrie are defined as
$$upper Barrie=(1+EMA(volatility)*upfactor)*closing  Price,$$
$$lower Barrie=(1+EMA(volatility)*lowerfactor)*closing Price,$$
where upfactor and lowerfactor are the predefined parameter of volatility sensitivity. This is clearly a path-dependent approach, as we need to determine whether the three barriers are touched at some point during the entire time interval. 
\begin{figure}[htb]
\centering
\includegraphics[scale=0.5]{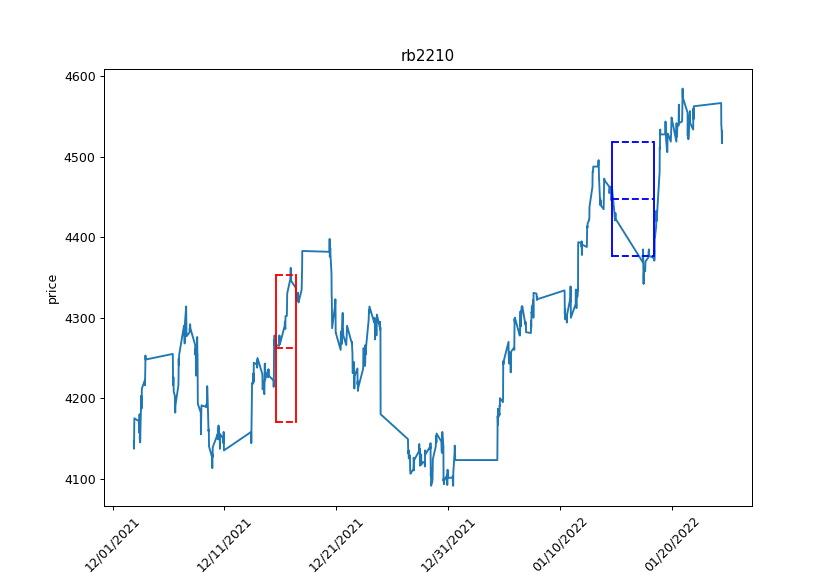}
\caption{Labeling chart with triple barrer method}
\label{fig-7}
\end{figure}
Figure \ref{fig-7} shows the configuration of the triple barrier method. As shown, the price range in red should be labelled 1, because the top barrier was first hit and this part of the chart is an uptrend; the price range in blue should be labelled -1, because the bottom barrier was first hit and this part of the chart is a downtrend.

\section{Automated Trading System}

We set up the trading system in four steps:
\begin{itemize}
    \item \textbf{Differencing:} We perform a Fractionally Differentiated Method on the price data to ensure a stationary time series.~~\par
    \item \textbf{Labeling:} We label the data with the Triple-Barrer Method.~~\par
    \item \textbf{Modeling:} We construct a neural network model for multi-class classification with Module class of pytorch.~~\par
     \item \textbf{Trading:} Using the model's predictions, we use the quantitative trading platform VeighNa to set up a trading system.
  \end{itemize}
The first three steps are used to train the neural network model and the fourth step is to do mainly backtest simulations to build trading strategies based on the predictive labels.

\subsection{Training neural network models}
The model is trained with three steps: the first step is to pre-process the data, the second step is to train the model, and the third step is to perform model validation. 

The foundation of any machine learning model is the data that is used to be trained and tested. Regarding the pre-processing of the data, we take the example of the rebar futures contract on the Shanghai Futures Exchange, the current main contract is the October expiry contract rb2210. Our objective is to construct an intraday trading strategy where we select 10 minute period price data which includes opening prices, closing prices, high prices, low  prices and volume. Firstly, we perform a stationarity transformations of the data. The original series of closing prices has an ADF statistic of -0.3387, at a $95\%$ confidence level, the test’s critical value is -2.8618, the closing price  is nonstationary. With the help of fractional order difference theory, the ADF statistic value of the closing price data obtained after $d=0.3$ order of difference is -3.049, which satisfies the stationarity, and the Pearson correlation coefficient of the differenced closing price data and the original closing price data reaches 0.988. Secondly, we apply the python based technical analysis package talib to calculate the technical analysis indicators, and we obtain a total of 81 indicators such as MACD, RSI, KDJ, Bollinger bands, etc.. These technical indicators characterise the transformation of price data in six ways: Overlap Studies, Momentum Indicators, Volume Indicators, Cycle Indicators, Price Transform, Volatility Indicators, etc. We then apply the Triple-Barrer Method to label the data. 

Before feeding the data into the model, we need to normalise each column of the sample feature data. The purpose of doing the normalisation is to eliminate the effect of variables with different scales on the learning of parameters, such as the different scales between price and volume, which may cause potential difficulties when training the network\cite{rai2019advanced}. We add a vector of 1's to the label vector $Y$ by changing -1, 0 and 1 to 0, 1 and 2 respectively. This is done just for one-hot encoding easiness, a process in machine learning where categorical variables are represented by dummy variables. Data is then randomly partitioned between training data (80$\%$) and test data (20$\%$). Training data serves to progressively fit the model, while test data serves to evaluate the model's performance\cite{rai2019advanced}, this is done in order to avoid over-fitting the model.

\begin{figure}[htp]
 \centering
\includegraphics[scale=0.05]{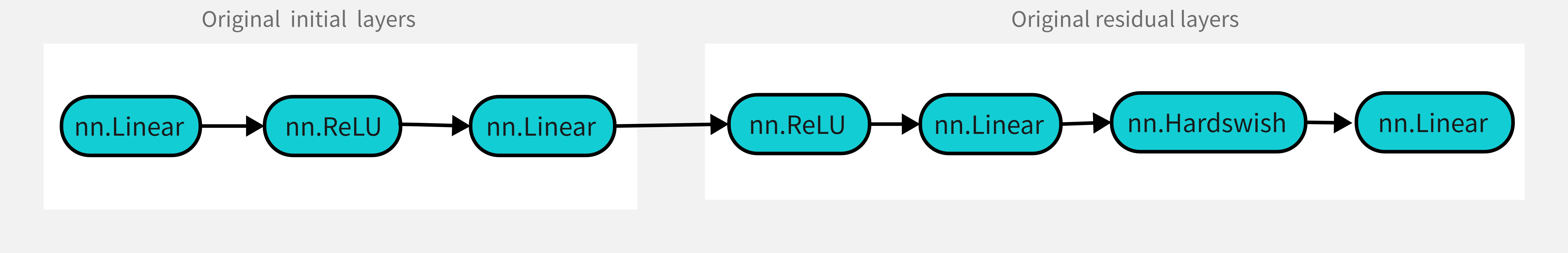}
\caption{Structural diagram of neural network models}
\label{fig-8}
\end{figure}

\begin{figure}[htb]
 \centering
\includegraphics[scale=0.088]{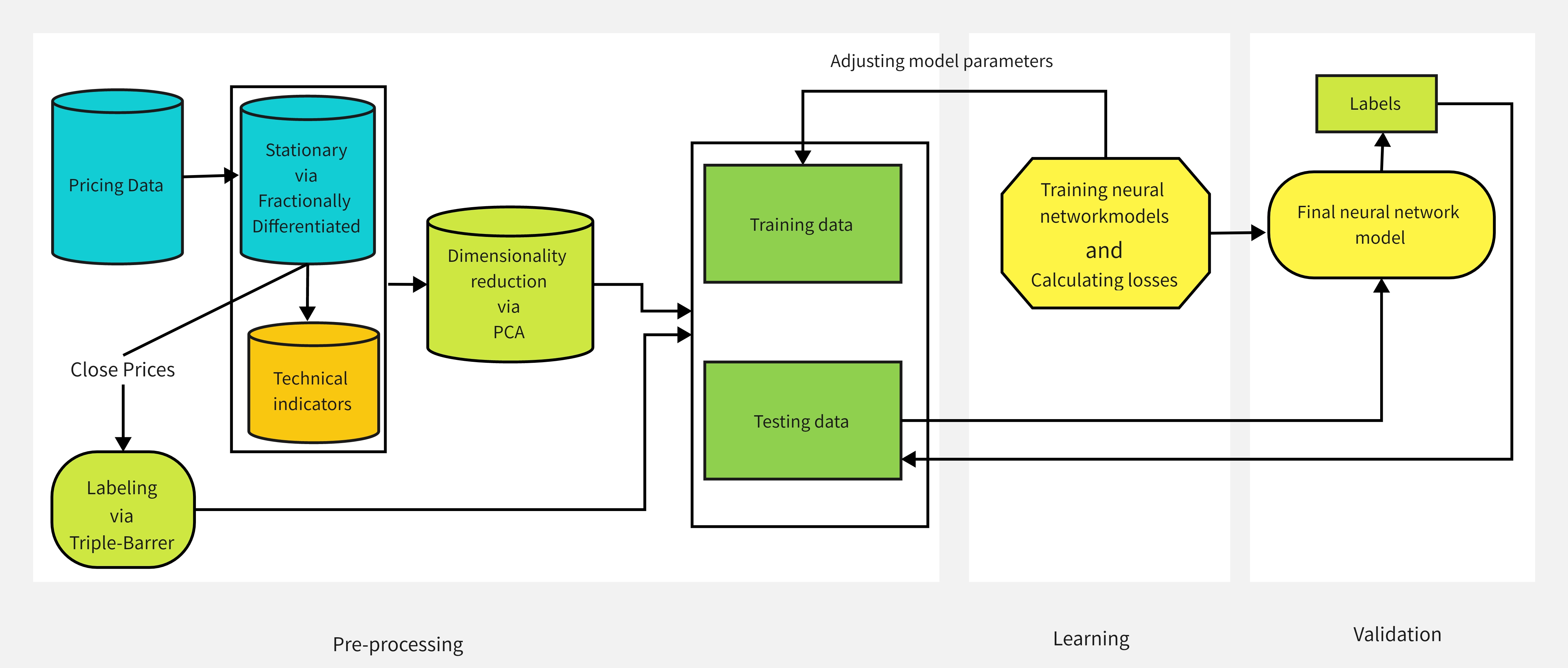}
\caption{Structural diagram of the supervised learning models}
\label{fig-9}
\end{figure}

Torch.nn is a modular interface designed specifically for neural networks, nn is built on top of Autograd and can be used to define and run neural networks. We build neural network models based on torch.nn. The original initial layers of the network consist of two Liner layers and a non-saturating activation function, the original residual layers of the network consist of two Liner layers and two unsaturated activation functions. The diagram below shows the structure of our network model. Figure \ref{fig-8} is the structural diagram of neural network models.
For our optimization, the loss function is chosen to be the cross-entropy loss function nn.CrossEntropyLoss(), and Adam was the chosen optimizer.  Figure \ref{fig-9} is the structural diagram of the supervised learning models. Our neural network model, trained on the data and corrected for parameters, yields the final predictive model.

\subsection{Trading Strategies}
To have a more tangible result, following the predicted labels with the final neural network model, we use VeighNa, a Python-based open source quantitative trading system, to build trading strategies. We design the trading rules as follows: when the position in futures is zero, we buy one lot of futures if the forecast label is 2; if the label is 0, we short one lot; if the label is 1, we do not operate. While placing the order, we set the take profit and stop loss conditions based on the volatility of the closing price. 

When holding a long futures order: 
\begin{itemize}
    \item \textbf{Take profit:} closing price$\times$(1+pa$\times$volatility).~~\par
    \item \textbf{Take stop:} closing price$\times$(1-pb$\times$volatility).   
  \end{itemize}

\begin{figure}[htb]
 \centering
\includegraphics[scale=0.088]{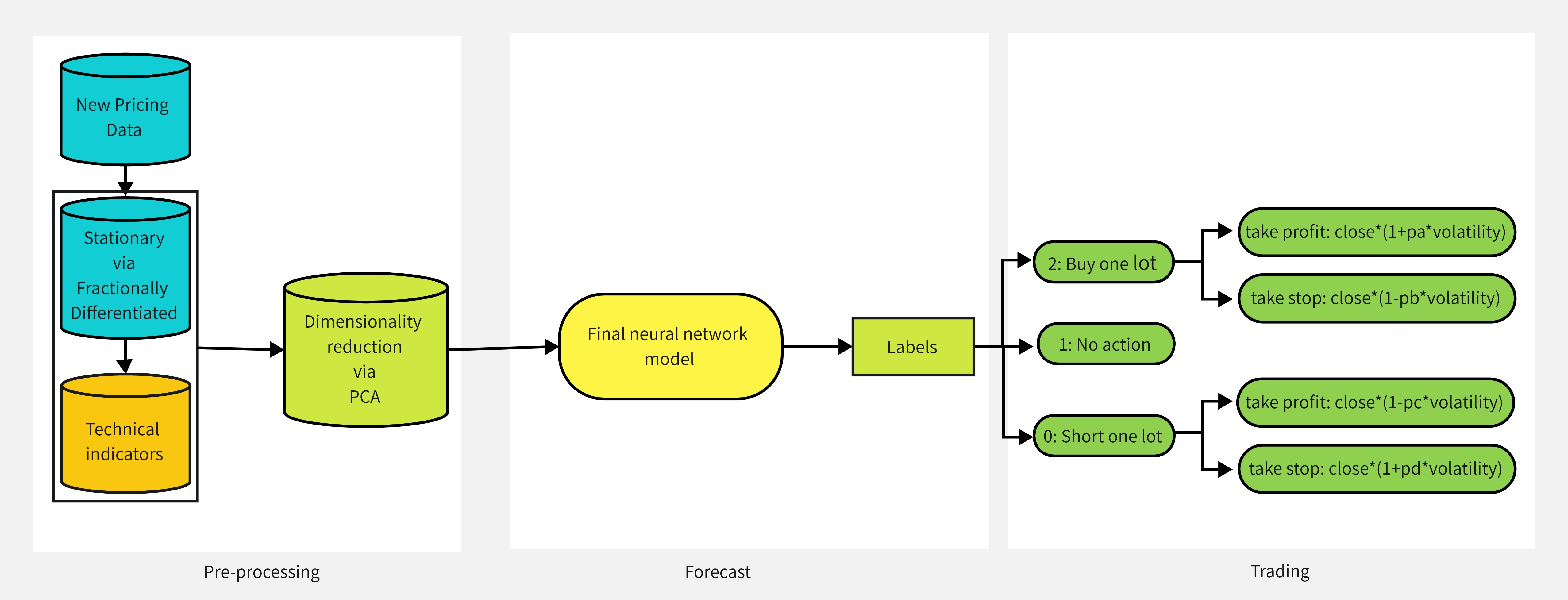}
\caption{Structural diagram of trading rules}
\label{fig-10}
\end{figure}

When holding a short futures order: 
\begin{itemize}
    \item \textbf{Take profit:} closing price$\times$(1-pc$\times$volatility).~~\par
    \item \textbf{Take stop:} closing price$\times$(1+pd$\times$volatility).   
  \end{itemize}
The hyperparameters pa, pb, pc, pd in the above trading rule are given multiplicative coefficients that we optimise in backtesting by a genetic algorithm nested in VeighNa to obtain the optimal solution. Figure \ref{fig-10} is the structural diagram of trading rules.

\subsection{Backtesting}
We chose the ten minute cycle data of the main rebar contract index for backtesting analysis. We found through simulation that data periods that are too short, due to too much noise in the data, the system will prompt frequent trades, which end up with poor trading results due to the presence of transaction costs and trading slippage. We wanted to construct an intraday trading strategy with few intraday trading opportunities if the period was too long.
The timeline for the trading data is from 15 December 2022 to 24 May 2022, with the first 80$\%$ of the data used for training and the second 20$\%$ for testing. Firstly, we synthesise the ten minute period data and then perform a fractional order difference transformation with $d = 0.3$, an enhanced Dickey-Fuller test on the differentiated series rejected the null hypothesis and concluded that no unit root existed, thus yielding a stationary series. By using the technical analysis package talib, we calculate the technical indicators, plus the price data, so that each of our sample data is an 86-dimensional vector, and with the help of Principal component analysis (PCA), the sample data is dimensioned down so that the input data is a 16-dimensional vector. When using the Triple barrier method to label data, we define a time limit window of 12 cycles and the multiplier for both the upper and lower multipliers is defined as 3. The hyperparameters in the trading strategy are optimised and determined as follows: [pa,pb,pc,pd]=[5,2,5,2]. 

\begin{figure}[htbp]
	\centering
	\begin{minipage}{0.75\linewidth}
		\centering
		\includegraphics[width=0.9\linewidth]{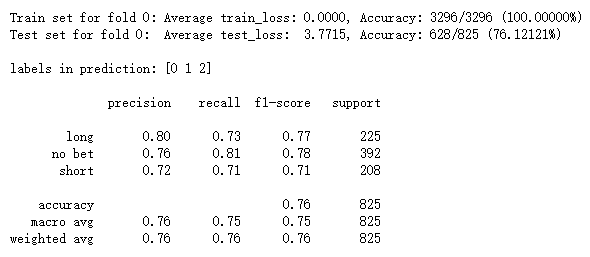}
		\caption{The classification report of test data}
		\label{fig-11}
	\end{minipage}
	\qquad
	
	\vskip 0.4cm
	\begin{minipage}{0.6\linewidth}
		\centering
		\includegraphics[width=0.9\linewidth]{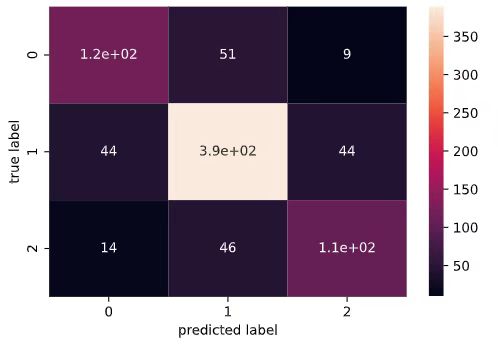}
		\caption{The confusion matrix of test data}
		\label{fig-12}
	\end{minipage}
\end{figure}

We use the classification model evaluation metrics Precision, Recall, and F1-score to characterize the predictive performance of the model on the test set. Our predicted outcome indicator results are divided into three categories: 0, 1, 2. For $i\in\{0,1,2\}$, Precision describes the number of indicators correctly predicted as indicator i, as a proportion of the total number of indicators predicted as indicator i; Recall describes the number of indicators correctly predicted as indicator i, as a proportion of the total number of indicators actually predicted as indicator i. The F1-score is the summed average of precision and recall; Support is the total number of samples in each classification test set; Macro avg is the average of all labeled results; Weighted avg is the weighted average of all labeled results. The Precision and Recall reports and illustration are shown in Figure \ref{fig-11}.
Here, we can see that the Precision of the proposed model is 80$\%$ for long, 76$\%$ for no bet and 72$\%$ for short. Figure \ref{fig-12} shows the confusion matrix of the model's predicted results on the test set. The generality of our classifier is largely guaranteed as our classifier has high single item Precision and Recall, which means that our classifier is sufficient to classify each segment of futures price movement.

\begin{figure}[htp]
 \centering
\includegraphics[scale=0.6]{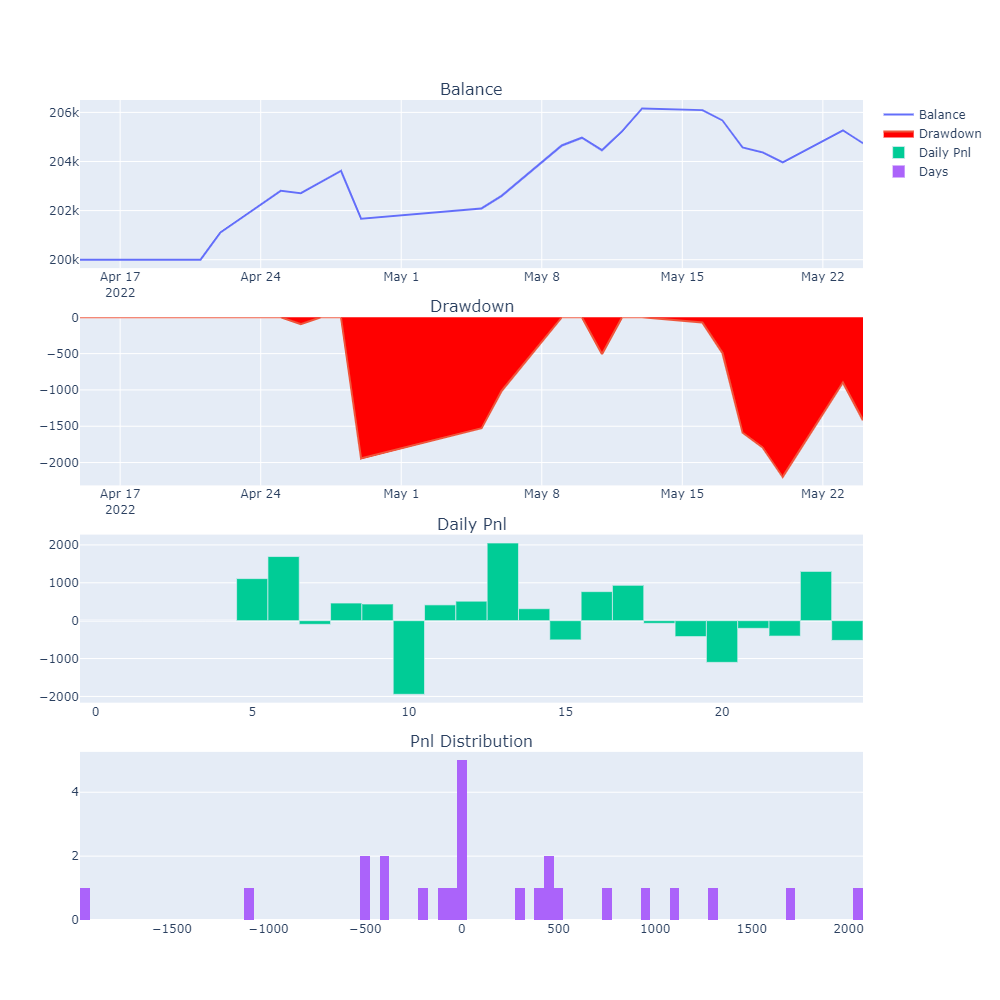}
\caption{Backtest analysis chart of the trading systems}
\label{fig-13}
\end{figure}

Figure \ref{fig-13}  shows the analysis of the backtest results. The backtesting configuration in CtaBacktester is as follows: we agreed on a backtest capital of RMB 200,000, 1 lot per open order, a commission rate of 0.00005 and a trade slippage of 1.0. The backtesting results show that out of 25 trading days, there were 11 profitable trading days and 9 losing trading days, with a maximum backtest of -1.07$\%$, an average of 2 trades per day and a total return of The total return was 2.37$\%$, with an annualised return of 22.77$\%$ and a Sharpe Ratio of 3.47.

\section{Conclusion}
This article builds a futures trading system based on a supervised learning model. We first transform the price data smoothly according to the theory of fractional order difference methods, then use the Triple Barrier Method method to label futures closing prices into three categories: Rising, falling and oscillating. We build a supervised learning model and design a trading strategy based on the prediction results of the model.The backtest results of the model prove the effectiveness of the trading system. Further research will be conducted in the following areas: firstly, the selection of data bars will be further optimised; secondly, the construction of the neural network model will be further investigated based on the structural characteristics of the sample data. Thirdly, the design of the trading strategy will be further optimised.

 \bibliography{myfile3}

\end{document}